\documentclass[aps,prl,twocolumn,groupedaddress,showpacs]{revtex4}
\usepackage{epsfig}
\usepackage[dvipsnames,usenames]{color}

\begin{document}

\title{Kondo screening and Magnetism at Interfaces}

\author{A. Euverte$^1$, F. H\'ebert$^1$, S. Chiesa$^2$,
  R.T.~Scalettar$^3$, G.G. Batrouni$^{1,4}$}

\affiliation{$^1$INLN, Universit\'e de Nice-Sophia Antipolis, CNRS;
1361 route des Lucioles, 06560 Valbonne, France }

\affiliation{$^2$ Department of Physics, College of William \& Mary,
Williamsburg, VA 23185, USA}

\affiliation{$^3$Physics Department, University of California, Davis,
California 95616, USA}

\affiliation{$^4$Institut Universitaire de France}

\begin{abstract}
The nature of magnetic order and transport properties near surfaces is
a topic of great current interest.  Here we model metal-insulator
interfaces with a multi-layer system governed by a tight-binding
Hamiltonian in which the interaction is non-zero on one set of
adjacent planes and zero on another. As the interface hybridization is
tuned, magnetic and metallic properties undergo an evolution that
reflects the competition between anti-ferromagnetism and (Kondo)
singlet formation in a scenario similar to that occurring in
heavy-fermion materials.  For a few-layer system at intermediate
hybridization, a Kondo insulating phase results where magnetic order
and conductivity are suppressed in all layers.  As more insulating
layers are added, magnetic order is restored in all correlated layers
except that at the interface. Residual signs of Kondo physics are
however evident in the bulk as a substantial reduction of the order
parameter in the 2-3 layers immediately adjacent to the interfacial
one. We find no signature of long range magnetic order in the metallic
layers.
\end{abstract}

\maketitle

%%%%%%%%%%%%%%%%%%%%%%%%%%%%%%%%%%%%%%%%%%%%%%%%%%%%%%%%%%%%%%%%%%
%\section{INTRODUCTION}
%%%%%%%%%%%%%%%%%%%%%%%%%%%%%%%%%%%%%%%%%%%%%%%%%%%%%%%%%%%%%%%%%%

Sufficiently strong electronic correlations can cause the formation of
an insulating phase at commensurate fillings.  In general, there is a
non-zero critical interaction strength required for this ``Mott
transition'', so that two uncoupled bands with different degrees of
correlation can coexist in metallic and insulating states. The
behavior of spectral functions and magnetic and superconducting
correlations, when interband hopping or interactions are turned on, is
a challenging theoretical problem.  The coupling could immediately
force both bands to be in the same (metallic or insulating) phase, or
coexistence might persist up to some critical degree of coupling
\cite{liebsch04,arita05}.

Closely related questions arise as clean interfaces between correlated
materials become accessible \cite{mannhart05}. Here the role of
different orbitals is played by the multiple layers.  It has been
suggested that it might be possible to ``engineer'' specific forms of
spectral functions at the interface by varying the materials
partnered, as well as design other properties arising from electronic
interactions \cite{mannhart10,freericks06,millis05}.  Experimental
realizations include tunable 2D electron gases in oxide
(SrTiO$_3$/LaAlO$_3$) heterostructures, control of magnetoresistance
at manganite interfaces \cite{hwang96}, novel magnetic properties at
boundaries between cuprate superconductors \cite{chakhalian06}, and
observation of magnetic proximity effect in Cu/CuO interfaces
\cite{munakata11}.

While the detailed chemistry of both multi-orbital and layered
materials is complex, an interesting starting point for studying the
qualitative properties of metal-insulator interfaces is provided by
the multi-layer Hubbard Hamiltonian.  In this model, electrons have
both intralayer and interlayer hopping, as well as layer-dependent
contact interactions.  The parameter space is large and in this paper
we focus on the simplest realization of the physics of a
metal-insulator interface in which all hybridizations are chosen to be
equal except the one at the interface; the corresponding Hamiltonian
is
\begin{eqnarray}
\label{Hamiltonian}
{\hat\mathcal H}&=&-t\!\!\sum_{\langle ij \rangle,  l, \sigma} \!\!
(c_{il\sigma}^\dagger c_{jl\sigma}^{\phantom{\dagger}} + 
{\rm h.c.})
-\mu \sum_{i,l,\sigma} n_{il\sigma}
\\
\nonumber
&&+ \sum_{i,l} U_{l} (n_{il\uparrow}-1/2)
(n_{il\downarrow}- 1/2)\\
&&-\!\!\! \sum_{i, \langle l l^\prime \rangle,  \sigma} \!\!
t_{ll^\prime}^{\phantom{\dagger}}
(c_{i l \sigma}^\dagger c_{i l^\prime\sigma}^{\phantom{\dagger}}
+ {\rm h.c.}).
\nonumber
\end{eqnarray}
Here $c_{i\,l\,\sigma}^\dagger(c_{i\,l\,\sigma}^{\phantom{\dagger}})$
are creation (destruction) operators for fermions of spin $\sigma$ at
site $i$ in layer $l$. Each layer is an $N$-site square lattice with a
contact interaction $U_l$ chosen to be non zero, $U_l=U$, on
``correlated" layers $l=1,2,3,\cdots$, and zero for an additional set
of ``metallic" layers ($l=-1,-2,-3, \cdots$). Layers are arranged in
order of increasing $l$ so that $l=\pm1$ label the layers at the
interface.  $t$ and $t_{ll^\prime}$ are the intra and interlayer
nearest-neighbor hybridizations.  $t_{ll^\prime}=t$ except at the
interface where it takes the value $t_{-1,1}=V$.  We consider the case
where $\mu=0$ which, as a consequence of particle-hole symmetry, makes
all layers half-filled, $\langle n_{il\sigma}\rangle=0.5 $.  Recent
studies on similar models have found induced magnetic order in the
metal \cite{sherman10} and quasi-particle penetration in paramagnetic
Mott insulators \cite{helmes08}.

Questions that arise in connection with Hamiltonian
(\ref{Hamiltonian}) can be seen as extensions to those typically asked
in the context of heavy-Fermion materials \cite{foot1} and concern, at
least at half-filling, the competition of magnetic order and screening
of local moments by conduction electrons.  Heavy-Fermion materials are
modeled by Hamiltonian (\ref{Hamiltonian}), a bilayer with $l=\pm 1$,
or by its strong coupling limit, the Kondo-Heisenberg lattice, where
charge fluctuations on the correlated layers are neglected.  The
fundamental issue we address here is how this competition is affected
as the two-layer case crosses over to the 3-dimensional bulk-to-bulk
interface.

At small and large interface hybridization, the system is
adiabatically connected to, respectively, the $V=0$ and $V=\infty$
limits.  At small $V$ the system is made up of magnetically ordered
layers weakly coupled to a metal.  At large $V$ the central bilayer
decouples and leaves the external layers either metallic ($l\le-2)$ or
insulating, with anti-ferromagnetic long range order ($l\ge +2$).

Our results indicate that, for a four-layer system, two interacting
and two metallic sheets, and at intermediate interfacial
hybridization, there exists an intervening phase where loss of
anti-ferromagnetic order is seen in both correlated layers, $+1$ and
$+2$, despite the fact that the latter is not in direct contact with
the metal.  We found that the electronic structure of the metal is
also profoundly affected and that the overall phase of the quad-layer
can be characterized as a Kondo insulator.  This is in contrast to our
other finding when the interaction region becomes thicker than the
metallic one: no loss of magnetic order is found in layers beyond the
one immediately adjacent to the metal, regardless of the hybridization
strength $V$.  In this case, upon increasing $V$, a direct transition
between the small and large $V$ regimes results.

%%%%%%%%%%%%%%%%%%%%%%%%%%%%%%%%%%%%%%%%%%%%%%%%%%%%%%%%%%%%%%%%%%
%\section{HAMILTONIAN AND METHODS}
%%%%%%%%%%%%%%%%%%%%%%%%%%%%%%%%%%%%%%%%%%%%%%%%%%%%%%%%%%%%%%%%%%

We addressed the physics of Hamiltonian (\ref{Hamiltonian}) using
determinant Quantum Monte Carlo (DQMC) \cite{blankenbecler81}, an
exact, finite-$T$ method for solving tight binding Hamiltonians on
finite lattices. As we limit our calculations to the perfectly
half-filled case, there is no sign problem at any temperature. Our
results are averaged over several independent simulations, and the
error bars correspond to the standard deviation of the mean.  The
imaginary-time step is set to $\Delta \tau = t/8$. We present results
for the in-plane anti-ferromagnetic structure factor,
\begin{eqnarray}
S^{\rm af}_{l}\equiv
\frac{1}{3N}\sum_{i,j}(-1)^{i+j}\big[2\langle\sigma^{x}_{il
}\sigma^{x}_{jl}\rangle+\langle\sigma^{z}_{il}\sigma^{z}_{jl}\rangle\big],
\end{eqnarray}
where $ \sigma^{x}_{il} = c_{il\uparrow}^{\dagger}
c_{il\downarrow}^{\phantom{\dagger}} + c_{il\downarrow}^{\dagger}
c_{il\uparrow}^{\phantom{\dagger}}$ and $ \sigma^{z}_{il} =
c_{il\uparrow}^{\dagger} c_{il\uparrow}^{\phantom{\dagger}} -
c_{il\downarrow}^{\dagger} c_{il\downarrow}^{\phantom{\dagger}},
\nonumber $ and the local layer dependent spectral function
$A_l(\omega)$, obtained by inverting the integral equation
\begin{eqnarray}
G_l(\tau) = \int_{-\infty}^{+\infty} d \, \omega \, 
\frac{e^{-\omega \tau}}{1+e^{-\beta \omega}} 
\, A_l(\omega)
\end{eqnarray}
via the maximum entropy method \cite{gubernatis91}.  $G_l(\tau) =
\sum_{i\sigma} \langle T\, c_{i l\sigma}(\tau)^{\phantom{\dagger}}
c_{i l\sigma}^{\dagger}(0) \, \rangle$ is the quantity directly
obtainable by DQMC, and is averaged over 4 boundary conditions
corresponding to setting the hopping at the boundary of each layer to
$\pm t$ \cite{gros92}.  We also study the in-layer electrical
conductivity $\sigma_l$ which is extracted from the current-current
correlation function
\begin{eqnarray}
 \Lambda_{xx,l}(k,\tau) = \sum_{i\in l} e^{ik\cdot i} \langle
 j_{x}(i,\tau)j_{x}(0,0) \rangle, 
\label{lambdaxx}
\end{eqnarray}
with $j_{x}(r,0) = it \sum_{\sigma}
(c^\dagger_{r+x,\sigma}c^{\phantom{\dagger}}_{r,\sigma} -
c^\dagger_{r+x,\sigma}c^{\phantom{\dagger}}_{r,\sigma})$.  We focused
on the intralayer contribution to $\Lambda_{xx,l}$ only, assuming that
this correctly characterizes the conductive property of each layer.
$\sigma_l$ is extracted using the approximate form of the
fluctuation-dissipation relation, valid at large $\beta$ and first
discussed in [\onlinecite{randeria92}],
\begin{eqnarray}
 \Lambda_{xx,l}(k=0,\tau=\beta/2) = \pi\sigma_l/\beta^2.
\end{eqnarray}

%%%%%%%%%%%%%%%%%%%%%%%%%%%%%%%%%%%%%%%%%%%%%%%%%%%%%%%%%%%%%%%%%%
%\section{RESULTS}
%%%%%%%%%%%%%%%%%%%%%%%%%%%%%%%%%%%%%%%%%%%%%%%%%%%%%%%%%%%%%%%%%%

\begin{figure}[ht]
\centerline{\epsfig{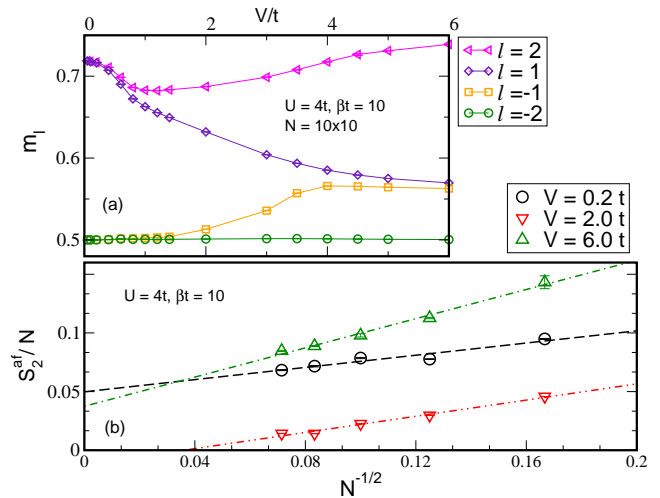}}
\caption{(color online) (a) $V$-dependence of local moments,
  $m_l=\sum_i \langle (\sigma^z_{il})^2\rangle/N$, on each layer, when
  two metallic layers are coupled to two correlated ones.  (b) Finite
  size scaling of in-plane structure factor $S^{\rm af}_2$ of the
  correlated layer ($l=2$) farthest from the interface.  For small
  $V$, there is long range order in the thermodynamic limit which
  vanishes for intermediate $V$ and is recovered for large $V$.
  $S^{\rm af}_2$ reaches its ground state value at $\beta t=10$.
\label{fig1}
}
\end{figure}

To gain initial quantitative understanding of the evolution of
magnetic properties as $V$ increases, we show, in the top panel of
Fig.~\ref{fig1}, the evolution of local moments in a system of two
metallic and two interacting sheets. There are three regimes, most
clearly evidenced by the behavior of the metallic layer at the
interface.  At $V\lesssim t$ local moments on layer $-1$ are
essentially identical to those of a non-interacting system.  In
$t\lesssim V \lesssim 4t$ the moments monotonically increase and they
saturate at $V\simeq 4t$.  The evolution of the other layers follows
naturally, with layer $+1$ merging with $-1$ at large $V$ in a phase
that can be best characterized as a band insulator made of weakly
interacting dimers. Layer $+2$ has the only non-monotonic evolution:
magnetism is first suppressed and then revived as the central dimer
phase gets increasingly stabilized.

We then use finite size scaling on $S^{\rm af}_2$ to investigate
whether order in layer $+2$ is lost in the regime where the moments
are most suppressed.  As shown in the lower panel of Fig. \ref{fig1},
$S^{\rm af}_2$ scales to a nonzero value for both small and large
interface hopping, $V$, when plotted against the inverse linear system
size $1/\sqrt{N}$.  For both these regimes, there is long range order
in the ground state in the thermodynamic limit \cite{huse88} as
expected from the behavior of the local moments.  However, in the
intermediate regime (starting at $V\ge t$, in rough correspondence to
where loss of anti-ferromagnetic order happens in the periodic
Anderson model \cite{vekic95} and the bilayer Hubbard model
\cite{scalettar94}) loss of magnetic order is clearly observed on
layer $+2$ as well.

\begin{figure}[t]
\centerline{
\epsfig{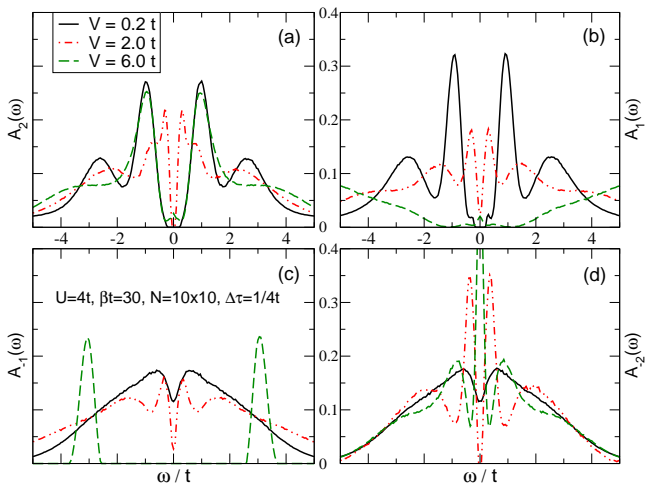}}
\caption{(color online) (a) Spectral function $A_2(\omega)$ in
  correlated layer $l=2$. The Slater gap present at small $V$ due to
  AF order vanishes at intermediate $V$ and reappears at larger $V$.
  (b) In correlated layer $l=1$, $A_1(\omega)$ resembles $A_2(\omega)$
  but unlike $l=2$ the large $V$ behavior is a broader gap associated
  with the singlet energy scale.  In both non-interacting layers
  $A_{-1}(\omega)$ (c) and $A_{-2}(\omega)$ (d) a gap opens as $V$
  increases. At large $V$, layer $-2$ recovers metallic properties
  while the singlet gap is visible in layer $-1$.
\label{fig2}
}
\end{figure}

The most likely candidate mechanism for such loss of order involves
Kondo screening in both interacting layers. As the formation of a
resonance in the single particle spectral density is one of the
hallmark of such process, we plot, in Fig.\ref{fig2}, the layer
dependent spectral density at $T=t/30$.  We found that, at $V=2t$,
both interacting layers are characterized by the presence of a Kondo
resonance. Due to the fact that we are focusing on a half-filled
system, the resonance is split as typically happens for Kondo
insulators.

\vskip0.1in

\begin{figure}[t]
\centerline{\epsfig{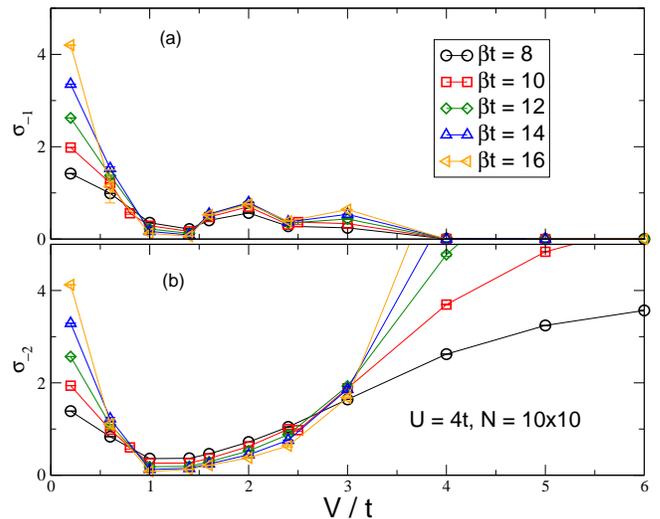}}
\caption{(color online) (a) In-plane conductivity $\sigma_{-1}$ in the
  metallic layer $l=-1$ as a function of $V$, at several inverse
  temperature $\beta$ values. At intermediate $V$ is small but
  non-zero, but vanishes at $V\geq4t$ due to dimer formation.  (b)
  In-plane conductivity $\sigma_{-2}$. This noninteracting layer
  becomes insulating for intermediate hybridization, $t\leq V\lesssim
  3t$ and then recovers when the pairs are fully formed and pinned at
  the interface.
\label{fig3} }
\end{figure}

We can gain further insight into the nature of this intermediate phase
by looking at the behavior of the non-interacting layers.  Figure
\ref{fig3} shows the conductivity in layers $l=-1$ and $l=-2$.  At
small hybridization, before the loss of magnetic order, the
conductivity increases as $T$ is lowered, showing these two sheets to
be metallic. In the intermediate regime $t \lesssim V \lesssim 3t$,
the conductivity is strongly suppressed in {\em both} layers, and our
inverse temperature results suggest that these layers become
insulating around $V=t$ in correspondence with the loss of magnetic
order.

From these results we can draw a few significant conclusions.  First,
that there is a Kondo proximity effect, as already observed using
dynamical mean-field theory on a similar model \cite{helmes08}.  The
novelty of our finding resides in our treatment of non-local
correlation. It allows for a proper description of the competition
between magnetic order and Kondo screening and shows the latter to be
effective in destroying order even on layers not directly coupled to
the metal.  Furthermore, states from both metallic layers participate
in the screening of local moments as best evidenced by the drop in
conductivity.  This situation is reminiscent of a long-range
resonating valence-bond state although it is unclear whether such a
description remains meaningful in the present context of itinerant
electrons.

The other kind of proximity effect that is expected in such systems is
due to the presence of the magnetically ordered layers at smaller
values of $V$.  This would seem a likely scenario, especially since
the metallic phase lives on layers with nested Fermi surfaces with
infinite, $T=0$, anti-ferromagnetic susceptibility.  Our calculations,
however, do not find any significant penetration of magnetic order in
the metallic layers.  Although we cannot exclude that an extremely
small order parameter might develop at low $T$ for some range of $V$,
our finding suggests that such an order would not survive the generic
scenario of a system with finite susceptibility.  This result appears
at odds with recent experiments \cite{munakata11} finding evidence for
an anti-ferromagnetic proximity effect.

\begin{figure}[t]
\centerline{\epsfig{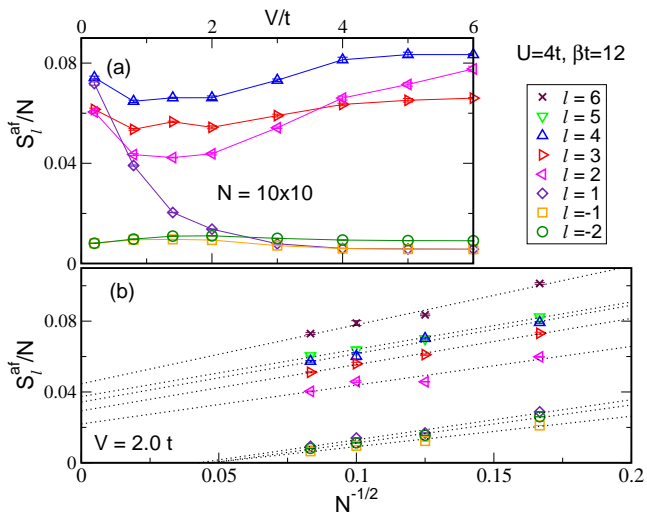}}
\caption{(color online) (a) In-plane structure factors $S_l^{\rm af}$
  as functions of $V$ for a system of four correlated layers coupled
  to two metallic ones.  At intermediate $V$ antiferromagnetic order
  is suppressed in every correlated layer. At large $V$ the order
  recovers in every correlated layer but the one at the interface.
  (b) Finite-size scaling of the in-plane structure factors $S_l^{\rm
    af}$ for a system of six correlated layers coupled to two metallic
  ones, at $V=2t$.  A systematic reduction of spin correlations is
  evident as the metallic interface is approached from the correlated
  side.  Long range magnetic order is completely destroyed in the
  correlated layer at the junction.
\label{fig4} }
\end{figure}

We now consider the question of how the competition of Kondo screening
and magnetic order is affected when additional interacting layers
$l=3,4,\cdots$ are present.  The resulting system can be thought of
describing the interface between a thin metallic film and a bulk
anti-ferromagnetic Mott insulator or as a heterostructure where the
insulating domains are substantially thicker than the metallic ones.
A scan of $S_l^{\rm af}$ for different values of $V$ on a 6-layer
cluster with 600 sites (top panel in Fig. \ref{fig4}) indicates that
the most likely parameter regime to observe loss of magnetic order in
layer 2 is for $V/t\in[1,2]$.

However, for $V=2t$ (lower panel of Fig. \ref{fig4}), we found that,
while $S^{\rm af}_1$ does not extrapolate to a non-zero value, the
same does not happen for layers located deeper into the interacting
material {\em i.e.} $l=2,3,4,5,6$ are all anti-ferromagnetically
ordered.  This revival of magnetic long-range order on layer $+2$ can
be interpreted as a magnetic proximity effect exerted by a bulk
anti-ferromagnet on those correlated layers subject to Kondo
screening.  Note that a clear suppression of the order parameter is
still observed for $l=2$ and several layers deeper in indicating
coexistence of Kondo screening and magnetism (the anomalously large
value at $l=6$ is a known phenomenon where surface magnetic
correlations are larger than the bulk and has been widely explored
experimentally \cite{sinkovic85}).

%%%%%%%%%%%%%%%%%%%%%%%%%%%%%%%%%%%%%%%%%%%%%%%%%%%%%%%%%%%%%%%%%%
%\section{CONCLUSIONS}
%%%%%%%%%%%%%%%%%%%%%%%%%%%%%%%%%%%%%%%%%%%%%%%%%%%%%%%%%%%%%%%%%%

In conclusion, we have presented results on a model of metal-insulator
interface, the multilayer Hubbard Hamiltonian.  Even within this
simplified tight-binding model, there are many possible choices of the
intralayer and interlayer hoppings.  The dominant feature of the
coupling of the metal and strongly interacting material is a
suppression of magnetic order on the correlated side.  We did not
observe the converse phenomenon, namely a significant penetration of
magnetism into the metal, as has been noted in
[\onlinecite{sherman10}].  It is possible this difference arises from
the lower value of the on-site interaction, $U/t=4$, used here,
compared to $U/t\approx17$ in [\onlinecite{sherman10}].  Such large
couplings are difficult to treat in DQMC.  The on-site interactions
studied here are relevant to the range of fitted values for a number
of interesting materials, e.g.~CuO \cite{bacci91}.

We showed that for thin insulating layers the Kondo effect embraces
correlated and metallic layers that are not in direct contact with
each other. Although such an extended Kondo insulating phase does not
require any fine tuning, it is not a phase that permeates a large
fraction of the ``phase space'' for such systems.  Here we have shown,
for instance, that forming Kondo singlets across multiple layers is a
process that can be defeated by magnetic proximity effect of layers
farther from the interface.

%\section*{ACKNOWLEDGMENTS}

This work was supported by: the CNRS-UC Davis EPOCAL LIA joint
research grant; by NSF-PIF-1005503 and DOE SSAAP DE-FG52-09NA29464; 
ARO Award W911NF0710576 with funds from the DARPA OLE Program; and Army
Research Office Grant 56693-PH.

%%%%%%%%%%%%%%%%%%%%%%%%%%%%%%%%%%%%%%%%%%%%%%%%%%%%%%%%%%%%%%%%%%%%%%%%
%%%
%%%%%     BIBLIOGRAPHY 
%%%%%%%%%%%%%%%%%%%%%%%%%%%%%%%%%%%%%%%%%%%%%%%%%%%%%%%%%%%%%%%%%%%%%%%%%%%


\begin{thebibliography}{50}
\bibitem{foot1}
Of course in the case where $l$ is interpreted as a band index the
physically interesting choice of $t_{ll'}$ would differ
from that chosen here, where only `adjacent' layers $l$ are hybridized.
Even more significantly, the Hamiltonian would likely contain more
general (Hund's Rule) interactions.
\end{thebibliography}
\end{document}